\documentclass[aps,amsmath,twocolumn,amssymb,floatfix,showpacs,superscriptaddress,bibliography]{revtex4-1}

\usepackage{graphicx}
\usepackage{dcolumn}
\usepackage{multirow}
\usepackage{booktabs}
\usepackage{bm,color}
\usepackage{braket}
\usepackage{amsmath,amssymb}
\usepackage[colorlinks,linkcolor=blue,hyperindex,CJKbookmarks]{hyperref}
\usepackage{epstopdf}
\usepackage{xcolor}
\usepackage{titlesec}

\newcommand{\blue}[1]{\textcolor{blue}{#1}}

\begin {document}
\title{\bf Electronic structure and unconventional non-linear response in double Weyl semimetal SrSi$_2$}
\author{Banasree Sadhukhan}
\email{banasree@kth.se}
\affiliation{ KTH Royal Institute of Technology, AlbaNova University Center, SE-10691 Stockholm, Sweden}
\affiliation{Leibniz Institute for Solid State and Materials Research, IFW Dresden, Helmholtzstr. 20, 01069 Dresden, Germany}
\author{Tanay Nag}
\email{tnag@physik.rwth-aachen.de}
\affiliation{Institut f\"ur Theorie der Statistischen Physik, RWTH Aachen University, 52056 Aachen, Germany}

\date{\today}

\begin{abstract}
Considering a non-centrosymmetric, non-magnetic double Weyl semimetal (WSM) SrSi$_2$,  we
investigate the electron and hole pockets in bulk Fermi surface behavior that enables us to
characterize the material as a type-I WSM.
We study the structural handedness of the material and correlate it with the distinct 
surface Fermi surface at two opposite surfaces following an energy evolution. The Fermi arc singlet becomes doublet with the onset of spin orbit coupling  that  is
in accordance with the topological charge of the Weyl Nodes (WNs). A finite energy separation between 
WNs of opposite chirality in SrSi$_2$ allows us to compute   circular photogalvanic effect (CPGE). 
Followed by the three band formula, we show that CPGE is only quantized for Fermi level chosen in the vicinity of 
WN residing at higher value of energy. Surprisingly, for the other WN of opposite chirality  in the  lower value of energy, CPGE is not found to be quantized. Such a behavior of CPGE is in complete contrast to the time reversal breaking WSM where CPGE is quantized to two opposite plateau depending on the topological charge of the activated WN. We further analyze our finding by examining the momentum  resolved CPGE. Finally we show that two band formula for CPGE is not able to capture the quantization that is apprehended by the three band formula.

\end{abstract}

\maketitle

\section{Introduction}

\par The concept of chirality, determined by the fact  whether an object is superimposable with its mirror image, is  present in various research fields from biology to  high energy physics. The  chiral crystals have a well defined structural handedness due to the lack of inversion,  mirror,  or other roto-inversion symmetries. Such symmetry breaking manifest themselves through  many fascinating properties e.g., optical activity \cite{PhysRevB.92.235205},  negative refraction \cite{pendry2004chiral},  unusual superconductivity \cite{carnicom2018tarh2b2}, quantized circular photogalvanic effect (CPGE) \cite{de2017quantized, flicker2018chiral, de2020difference, chan2017photocurrents},  gyrotropic magnetic effect \cite{zhong2016gyrotropic},  unusual phonon dynamics, chiral magneto-electric effects \cite{rikken2001electrical, morimoto2016chiral}, magnetic Skyrmions \cite{bogdanov1994thermodynamically} and many more.  The structural handedness thus imprints its effect in topological responses for chiral semimetals. 
The fourfold degeneracy of the linear band touching
in Dirac semimetals (DSMs) such as Cd$_3$As$_2$ and Na$_3$Bi, is broken by either time reversal or inversion symmetry leading to  Weyl semimetals (WSMs) with isolated twofold non-trivial band crossings  \cite{jenkins20163d, borisenko2014experimental,  weyl1929elektron}.

In this process, WSM is found to exhibit Weyl nodes (WNs), protected by a certain crystalline symmetries,
that act as  monopoles
or anti-monopoles of Berry curvature
in momentum space, with integer topological charge $n$  \cite{RevModPhys.90.015001,  nagaosa2010anomalous,  xiao2010berry,yan2017topological, armitage2018weyl}. 
On the other hand, 
DSMs have  net vanishing topological charge $n=0$ \cite{jenkins20163d, borisenko2014experimental,  weyl1929elektron}. 
As compared to the conventional
WSMs with $n = 1$ \cite{xu2015discovery,  lv2015observation,  lv2015experimental}, the multi-WSMs have higher topological charge $n>1$ with the crystalline symmetries bounding its
maximum value to three \cite{xu2011chern,  fang2012multi,  yang2014classification}.
The double-WSM (triple-WSM) show linear dispersion along one symmetry direction and quadratic (cubic) energy dispersion relations in remaining two directions respectively.
We note that higher topological charge is also observed for multifold band crossing  with integer spin in topological chiral crystals such as, the transition metal mono-silicides MSi (M = Co, Mn, Fe, Rh) \cite{schroter2019chiral, changdar2020electronic, ni2020giant,ni2020linear, le2020ab}. 
 The Fermi arc surface states and chiral anomaly induced negative magnetoresistance directly reflect the topological nature of WSM through its transport signatures \cite{zyuzin2012topological,  son2013chiral}.

\par Apart from first order electromagnetic and thermal responses \cite{yang2011quantum,   son2013chiral,  sharma2016nernst,  chen2016thermoelectric,  hirschberger2016chiral,  zhang2018strong,   nag2020magneto,   hirschberger2016chiral,  watzman2018dirac}, WSMs are further studied in the context of second order  transport response namely,   
CPGE. The WSMs  are found to exhibit quantized CPGE response where  direction of DC photocurrent depends on the helicity of the absorbed circularly polarized photons \cite{young2012first,  sipe2000second,  morimoto2016topological}.  
The  optical transitions near a WN plays an important role in quantization of CPGE serving as a
direct experimental probe to measure the Chern numbers in topological WSMs  \cite{yao2020observation,  sessi2020handedness}.   It is noteworthy that the breaking of inversion symmetry guarantees a finite CPGE. To be precise,  the breaking of inversion symmetry (while preserving mirror symmetry) and tilting of the Weyl cones in  non-centrosymmetric TaAs
family cause a giant non-zero and non-quantized CPGE response  \cite{ zhang2018photogalvanic,  de2020difference,  wu2017giant,  ma2017direct,  sun2017circular,  osterhoudt2019colossal,  lim2018temperature,  ji2019spatially}. On the other hand, 
non-degenerate WNs  can only result in a quantized CPGE referring to chiral nature of the underlying system where all mirror symmetries are broken \cite{de2017quantized,chang2018topological,  chang2018topological,  zhong2016gyrotropic}. More importantly, it has been shown  considering a time reversal symmetry invariant WSM lattice model that quantization in CPGE is substantially different from a time reversal broken WSM \cite{sadhukhan2020role}.

The double WSM state has been theoretically predicted recently in non-centrosymmetric SrSi$_2$ that preserves time reversal symmetry
\cite{huang2016new, singh2018tunable}.   From the technological perspective, in general, non-centrosymmetric WSMs can be useful in designing  the high-efficiency solar cells \cite{zhang2019switchable,zhang2018photogalvanic}.  
A finite energy separation between two WNs with opposite topological charge makes SrSi$_2$ an ideal material to study the CPGE. Hence a natural  question arises that how does  CPGE behave  in such a non-magnetic  double WSM. Our quest is indeed genuine due to  the fact that  SrSi$_2$
is categorically different from
TaAs WSM family as far as their electronic structures are concerned. At the same time, our study is  equally
relevant in the context of possible device application such as solar cell and experimental realizations.

In this work, we find  SrSi$_2$ 
has a structural handedness with enantiomeric properties. This is manifested in the  surface Fermi surface (SFS) profile where Fermi arc exhibits inverted structure in  two opposite surfaces $(001)$ and $(00\bar1)$.   The bulk Fermi surface, on the other hand, bears the information of electron and hole pocket referring to SrSi$_2$ as a type-I WSM.  We observe a quantized CPGE response in SrSi$_2$  when Fermi level is kept in the vicinity of the WN at higher energy while there is  no quantization  for other WN at lower energy. Our finding is in accordance with the fact that CPGE behaves qualitatively distinct manner time reversal invariant WSM as compared to time reversal broken WSM \cite{sadhukhan2020role}. 
We examine the evolution of CPGE with Fermi level considering spin orbit coupling (SOC) into the calculation and compare it to the bare calculation with  generalized gradient approximation (GGA) only.
Both the findings match qualitatively with each other while topological charge is found to be $n=2$ ($n=1$) followed by GGA+SOC (GGA) calculations.    
We anchor our findings with the momentum resolved analysis of CPGE. Furthermore, we check that two band formula for CPGE is unable predict the quantization that is correctly described by the three band formula. In short, our study uncovers various interesting non-linear optical response properties of SrSi$_2$ besides its implications based on structural properties.

The organization of the paper is as follows.  In Sec.~\ref{sec_II}, we discuss the crystal structure, electronic band structure, position of the WNs and other other bulk properties(Fermi surface, Berry curvature). In Sec.~\ref{sec_III}, we  present the surface properties such as surface Fermi surface, spectral energy distribution and Fermi arcs,  employing the slab geometry. We next 
describe the framework of  CPGE and the associated results in Sec.~\ref{sec_IV}. 
We provide the analysis on two band formula alongside there. We discuss the necessary computational details within the individual sections accordingly.
Finally, in Sec.~\ref{conclusion},  we end by the  conclusion and outlook.

\section{\color{blue}{ Topological characterization}}
\label{sec_II}

{\centering
\begin{figure}[ht]
\includegraphics[width=0.5\textwidth,angle=0]{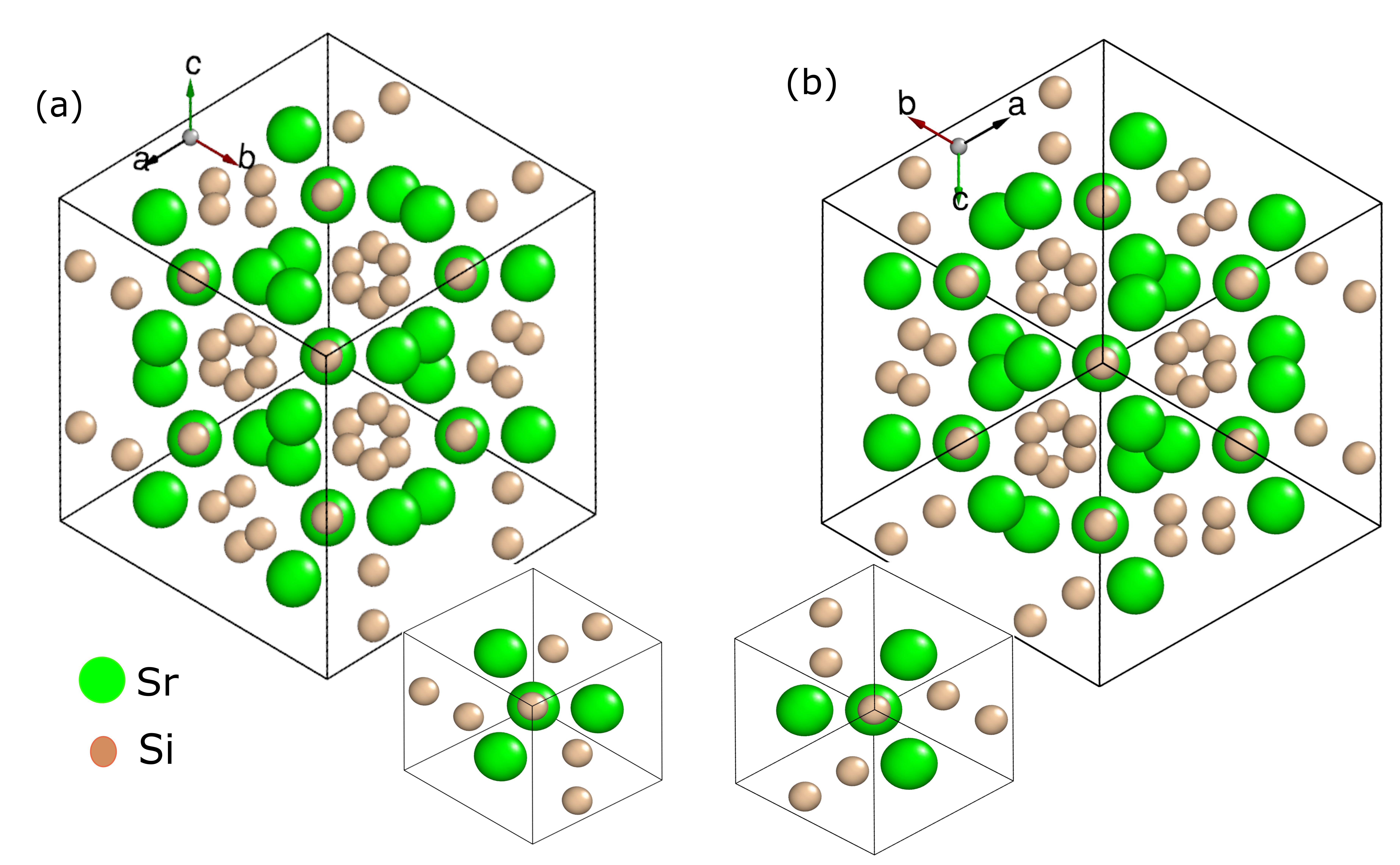}
\caption{ (a) Crystal structure of SrSi$_2$ showing the $(111)$  view of two enantiomers for single (lower) and double (upper) the unit cell.  (b) The crystal structure viewed from  $(\bar 1 \bar 1 \bar 1)$ which is the mirror ($a\to b$, $b \to a$ and $c\to c$) + flipped image ($a\to -b$, $b \to -a$ and $c\to -c$) of (a). 
Structural chirality generates a distinct handedness under a mirror operation referring to the enantiomeric property of the material. 
\label{fig:stuc}}
\end{figure}}

 \begin{figure*}[ht]
\hskip 0.9cm\includegraphics[width=1.01\textwidth,angle=0]{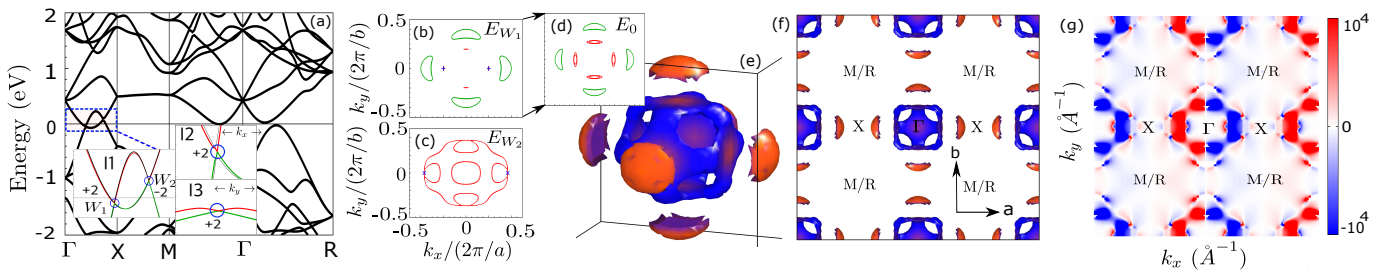}
\caption{(a) Bulk band structure of SrSi$_2$ from GGA+SOC calculation.  The energies of the double-WNs are   at $E_{W_1}=-26.3$ meV  and $E_{W_2}=88.2$ meV associated with topological charge $n=+2$ and $n=-2$, respectively as depicted in inset I1.  The insets I2 and I3 demonstrate the linear and quadratic band dispersion, around the WN of chirality $+2$, along $k_x$ and $k_y$, respectively. 
The bulk isoenergy surface for $k_z=0$ cut at $E_f= E_{W_1}$ (b), $E_f= E_{W_2}$ (c) and $E_f= 0$ (d). The Weyl points with positive and negative chiralities are marked by $+$ and $\times$, respectively.  
(e) Three dimensional  Fermi surface for a single unit cell and it's  projection (f) on the $k_x$-$k_y$ plane by doubling the unit cell.  (g) The $k_z$ averaged Berry curvature projected on the $k_x$-$k_y$ plane.
\label{fig:band}}
\end{figure*}

\par SrSi$_2$, crystallizing in a cubic Bravais lattice, has the chiral space group $P4_{3} 32$ ($\#$ $212$ ) with the lattice constant $
6.563 \AA$.  The unit cell contains four strontium (Sr) atoms and eight silicon (Si) atoms, which occupy the Wyckoff positions 4a and 8c, respectively.    The orientation of the atomic positions  responsible for handedness in SrSi$_2$.  Under a mirror operation, these two structural view in $(111)$ and $(\bar{1}\bar{1}\bar{1})$ direction, as shown in  Fig.~\ref{fig:stuc} (a)-(b), reverse their handedness. This fact can be used to distinguish the two enantiomers of SrSi$_2$ crystal.   Due to this unique chiral crystal structure,  SrSi$_2$ lacks both mirror and inversion symmetries, but has $C_2$, $C_3$ and $C_4$ rotational symmetries. Since SrSi$_2$ is a nonmagnetic system,  respecting the  time-reversal symmetry,  the absence of inversion symmetry is fundamental for realizing a WSM phase
with four WNs.

 \begin{table}[b!]
    \small
    \caption{Positions, Chern numbers, and energies of the WNs
    with SOC (W$_{1,2}$) and without SOC (V$_{1,2}$).}
    \begin{tabular*}{0.45\textwidth}{ p{1.0cm} p{4.0 cm} p {1.0 cm} p{1.3 cm} }
    \hline\hline
        WP  & Position [($k_x$,$k_y$,$k_z$) ]  & $ C $ & $E$ (meV)  \\
        &  in $(\frac{2\pi}{a}, \frac{2\pi}{b},\frac{2\pi}{c})$ &  &  \\
    \hline
     V$_1$ & ($\pm 0.2001, 0, 0$) & $+ 1$ & $-34.4$  \\
     V$_2$  & ($\pm  0.3691, 0, 0$) & $- 1$ & $\,\,\,\,82.3$ \\
    \hline
      W$_1$ & ($\pm  0.2003, 0, 0$) & $+ 2$ & $-26.3$  \\
     W$_2$ & ($\pm  0.3696, 0, 0$) & $- 2$ & $\,\,\,\,88.2$ \\   
        \hline
    \end{tabular*}
    \label{table:wp}
\end{table}

We study the electronic structure in SrSi$_2$ using DFT calculations. The Density functional theory (DFT) calculations are based on GGA with the ${\bm k}$-mesh 32$\times$32$\times$32 within the full-potential local-orbital (FPLO) code \cite{fplo_web}. The band structures from GGA+SOC along the high symmetry direction in Brillouin zone (BZ) are shown in the Fig.~\ref{fig:band} (a). We observe the band crossing between the highest occupied valence bands and the lowest unoccupied conduction along the $\Gamma-X$ direction. The WNs appear at $-34.4$ meV (V$_1$) and $82.3$ meV (V$_2$) with  topological charge $n=+1$ and $n=-1$, respectively, without SOC.  Under inclusion of SOC, the single WNs change their dispersion to form the double WNs in SrSi$_2$. These WNs appear at $-26.3$ meV (W$_1$) and $88.2$ meV  (W$_2$)  with  topological charge $n=+2$ and $n=-2$ respectively as shown in table \ref{table:wp}.

The band shown in green (inset Fig. \ref{fig:band}(a)) forms the top of the valence band and gives rise to two nested hole pockets along the line $\Gamma-X$.  Similarly, the band shown in red (inset I1 in Fig. \ref{fig:band}(a)) forms the  lowest conduction band which results two nested, closed electron pocket centering around the $\Gamma$ along the direction  $M-\Gamma-X$.
The linear and quadratic nature of the band dispersion along $k_x$ and $k_y$ are clearly shown in insets I2 and I3 while  focusing on the WN of chirality $+2$. 
Importantly, the WNs of double WSM SrSi$_2$ in presence of SOC are protected by time reversal and $C_2$ rotation symmetries while the inversion and mirror symmetries are already broken. Interestingly, in absence of SOC, the
quadratic dispersion turns into linear.  The size of the electron pockets  are larger than the hole pockets.
The conventional type-I WSMs are characterized by shrinking of the Fermi surface to a point at the WN energy. The simplest Fermi surface of such a WSM would consist of only two such points.
The cut of bulk three dimension Fermi surface for $k_z=0$ plane at different  energies  $E_f=E_{W_1}$, $0$, and $E_{W_2}$  are depicted in Fig.~\ref{fig:band} (b), (c) and (d), respectively. There exists only  hole (electron)  pocket for  $E_f=E_{W_1}$ ($E_f=E_{W_2}$) while Fermi surface hosts both electron and hole pockets simultaneously for any energy $E_{W_1}<E_f<E_{W_2}$. The energy separation between these single and double-WNs are $116.7$ and $114.5$ meV respectively resulting in  interesting chirality related phenomena namely, quantized CPGE response in  SrSi$_2$ which we describe in the subsequent section.

\par SrSi$_2$ exhibits a complicated and  nested bulk 3D Fermi surface (see Fig.~\ref{fig:band}(e)) whose projection in $k_x-k_y$ plane with the doubling of unit cell along with the high symmetry points of BZ is shown in Fig.~\ref{fig:band} (f).    We show the $k_z$ averaged Berry curvature in the $k_x-k_y$ plane, as depicted in Fig. \ref{fig:band} (f), by summing over all the occupied valence bands till  $E_f=0$. The sink ($C=-2$) and source ($C=2$) of the Berry flux that are marked by blue and red colors respectively in the Fig.~\ref{fig:band} (g).  The Berry curvature has the negative (positive) flux in the hole pockets along $\Gamma-X$ ($\Gamma-\bar{X}$) and electron pockets along $M-\Gamma-R$ ($\bar{M}-\Gamma-\bar{R}$) respectively.

\section  {\color{blue}{Topological surface states}}
\label{sec_III}

\begin{figure*}[ht]
\hskip -0.9 cm 
\includegraphics[width=1.05\textwidth,angle=0]{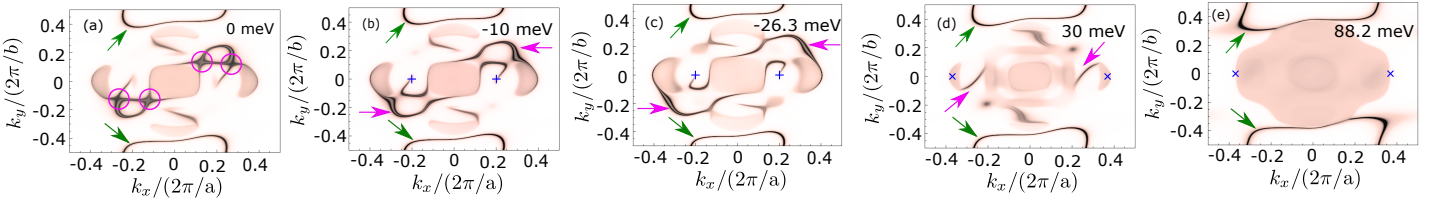}
\caption{ Surface Fermi surface of $(001)$-surface from GGA+SOC at (a) $E_f = 0$ ,  (b) $E_f = -10$ meV,  (c) $E_f = -26.3$ meV,  (d) $E_f = 30$ meV,  (e) $E_f = 88.2$ meV.  This shows the gradual formation of Fermi arcs i.e., non-trivial surface states (trivial surface states) with  the variation of $E_f$ marked by pink (green) arrows. }
\label{fig:fssd}
\end{figure*}

Figure \ref{fig:fssd} shows the evolution of the SFS mapping in the $(001)$-surface with decreasing (increasing) the energy from Fermi level $E_f=0$ meV to the  WN energy $E_f=E_{W_1}$ ($E_f=E_{W_2}$ ) using PYFPLO \cite{fplo_web} module as implemented within FPLO via Green’s function techniques.  We obtain a tight-binding Hamiltonian with 136 bands by projecting the Bloch wave functions onto Wannier functions. Here we consider Sr- 4d, 5s and 5p; Si- 3s and 3p orbitals within the energy range $-11.6$ eV to $10.0$ eV.  Here we will study the  gradual formation of Fermi arcs considering semi-infinite slab geometry.  The projection of big electron pocket appears around the $\Gamma$ point and hole pocket appears along $\Gamma-X$ direction at $E_f=0$ (see Fig.~\ref{fig:fssd} (a)). The surface states are found on the right (left) side for $k_x>0$ ($k_x<0$) in $(001)$-surface.   At $E_f=0$,  two Fermi arcs are connected around $k_x \approx \pm0.1$ and $k_x \approx \pm0.3$ (marked by circle in Fig. \ref{fig:fssd} (a)) in $(001)$-surface.   With the decreasing energy from $E_f=0$ to $E_f=-10$ meV,  the electron pockets around $\Gamma$ gradually shrinks and the hole pocket along $\Gamma-X$ increases as shown in Fig. \ref{fig:fssd} (b).  The Fermi arcs get splitted at the point around $k_x \approx \pm0.1$ and $k_x \approx \pm0.3$ and start to form the long $S$-shaped  Fermi arcs.  Finally at $E_f = E_{W_1} =-26.3$ meV,  the tail of the  long $S$-shaped Fermi arcs touch the WNs (see fig. \ref{fig:fssd} (c)).  With increasing the Fermi level from $E_f=0$ to $E_f = 30$ meV,  the electron pocket increases and hole pocket decreases (see fig. \ref{fig:fssd} (d)).  The Fermi arcs is gradually going into the  projected bulk band structure marked by arrows in Fig. \ref{fig:fssd} (d).  At $E_f =  E_{˝W_2} = 88.2$ meV,  the Fermi arcs touching the WNs around $k_x \approx \pm0.37$ are submerged into the bulk projected band structure as shown in Fig. \ref{fig:fssd} (e).  Additionally,  some trivial surface states running parallel to $|k_x|$ appear and forms a closed loop.  With increasing the Fermi level from $E_f=0$,  this trivial surface states touch the electron pockets at $E_f=88.2$ meV.  The same features in the  SFS mapping are also observed for GGA with a singlet Fermi arc  instead of Fermi arc doublet.

\begin{figure}[ht]
\includegraphics[width=0.5\textwidth,angle=0]{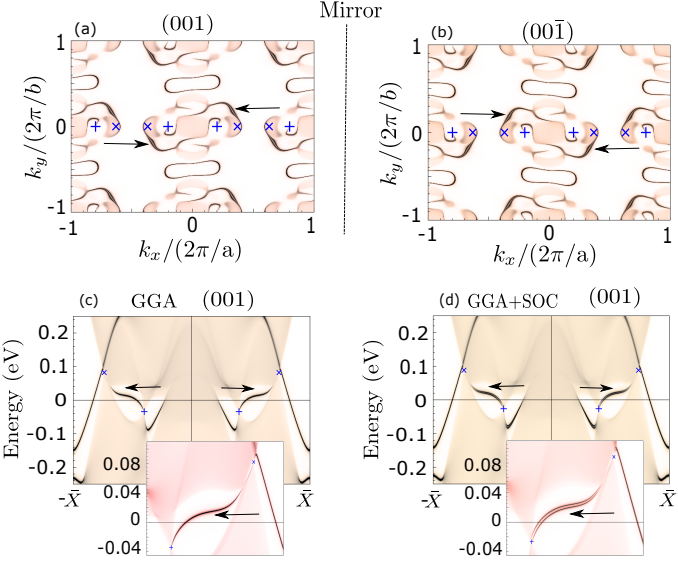}
\caption{  The surface Fermi surface of $(001)$-surface (a) and $(00\bar{1})$-surface (b) at $E_f=E_{W_1}$ respectively.  The long ``S"-shaped Fermi arcs doublet,  marked by arrows, connecting WNs with opposite chirality are found. The surface Fermi surface profiles on  $(00\bar{1})$-surface and $(001)$-surface are related by the  mirror reflection but not superimposable on each other.    The energy dispersion curve on the $(001)$-surface along $\frac{2\pi}{a}(k_x,{k_y}^{WP},0)$ from  GGA (c) and GGA+SOC (d) calculations.  The long chiral Fermi arcs singlet and doublet are clearly visible for GGA and GGA+SOC and marked by arrows.
\label{fig:fssd-edc}} 
\end{figure}

\par To further investigate the nature of the Fermi arc,  we calculate the  SFS states of $(001)$- and  $(00\bar{1})$-surface respectively
as shown in the Fig.~ \ref{fig:fssd-edc} (a)-(b) at  the WN energy $E_f=E_{W_2}=-26.3$ meV.    Two pair of WNs appear with opposite chirality and are marked by plus (cross) with the topological charge $n=+2$ ($n=-2$). The non-centrosymmetric compound SrSi$_2$ has two distinct surfaces for $(001)$ and $(00\bar{1})$ due to lack of mirror symmetry (see Fig.~\ref{fig:stuc} (a) and (b)). This causes the  SFS profile to possess a handedness that we describe below.  We observe a Fermi surface map with a large ``$S$"-shaped Fermi arcs in the full BZ connecting the two WNs of opposite chirality in the $(001)$  surface. Interestingly, the inverted ``$S$"-shaped Fermi arcs are observed in the $(00\bar 1)$  surface. 
Hence, the orientation of connecting pattern of Fermi arcs becomes reversed in $(001)$- and $(00\bar{1})$-surface. This property of Fermi surface is related to the enantiomer structure of the material where  mirror symmetric counterpart can not be 
superimposed with the parent structure.  We show in Fig.~\ref{fig:stuc} that the two enantiomers of SrSi$_2$ can be distinguished by the handedness of the crystal which is formed by their Sr/Si atoms along the $(111)$ direction.  Such a connection between Fermi arcs and crystalline handedness can lead to future studies on transports of chiral topological semimetal and their connection with the structural handedness.  Here the Fermi arcs are doubly splitted related to its magnitude of the topological charge $n=2$ and marked by arrows in  both the Fig.~\ref{fig:fssd-edc} (a)-(b).

To further elucidate the topological structure of the Fermi arcs,  we also calculated the energy dispersion curve  of the semi-infinite slab for $(001)$-surface the along the $\frac{2\pi}{a}(k_x,k^{WP}_y,0)$ from both GGA and GGA+SOC as shown in the Fig. \ref{fig:fssd-edc}.  The Fermi arcs in both cases are formed by a band which connects the top of the valence bands around $k_x=0.3$ to the bottom of the conduction around $k_x=0.2$ respectively.  The Fermi arcs are indicated by the arrows in the Fig. \ref{fig:fssd-edc} (c)-(d). The singlet and doublet  structures of Fermi arc
are clearly visible in GGA and GGA+SOC in accordance with the  the  magnitude of topological charges. The Fermi arc connecting WNs with $n=\pm2$ has positive (negative) chirality for $ \Gamma \to  X$ ($ \Gamma \to -X$)-direction.    This is in stark contrast to   the magnetic WSMs where a single Fermi arc connecting two WN of opposite topological charge is observed. The existence of such chiral Fermi arcs for time reversal symmetry broken WSM might lead to distinct transport signatures  as compared to the time  reversal symmetry invariant WSM. We also note that 
the energy dispersion in $(00\bar1)$-surface can be obtained by mirror reflection on that of the in $(001)$-surface. Hence the energy  dispersion in $(00\bar1)$ and  $(001)$-surfaces are superimposable with each other.   The energy dispersion profile 
follows a achiral pattern in the above surfaces where surprisingly, non-superimposable chiral nature of  SFS profiles are visible.

\section{\color{blue}{  Second order CPGE response}}
\label{sec_IV}

The CPGE injection current is a second order optical response when the system is irradiated with the circularly polarized light. It is defined as
\begin{equation}
\dfrac{d J_a}{dt} = \beta_{ab}(\omega) \left[\mathbf{E}(\omega)\times \mathbf{E}^{*}(\omega)\right]_{b},
\label{injection}
\end{equation}
where $\mathbf{E}(\omega)=\mathbf{E}^{*}(-\omega)$ is the circularly polarized electric field of frequency $\omega$, $a$ and $b$ indices are the direction of current $J_a$ and circular polarized light field respectively. Considering the relaxation time approximation and momentum independent relaxation time $\tau$ \cite{PhysRevB.88.104412,  lundgren2014thermoelectric,  ni2020giant},  we can introduce the 
broadening parameter  $ \Gamma = \hbar/ \tau$. The 
conductivity $\sigma_{ab}^c$ ($a,b,c = x,y,z$) is a third rank tensor representing the photocurrent $J_c$ generated by electric fields $E_a$ and $E_b$: $J_c=\sigma_{ab}^c {{E}^*_a}{E}_b$. We note that photoconductivity tensor $\sigma_{ab}^c$ and CPGE tensor $\beta_{cc}$ are essentially related by relation $\sigma_{ab}^c=\tau \beta_{cc}$.
From quadratic response theory,  the  photoconductivity reads as \cite{kraut1979anomalous,  von1981theory,  kristoffel1980some,  zhang2018photogalvanic}: 
\begin{eqnarray}
     &\sigma^c_{ab}(\omega) &= \frac{e^3}{ \omega^2} {\rm Re} \bigg\{ \phi_{ab}
                        \sum_{\Omega=\pm \omega} \sum_{l,m,n} \int_{BZ} \frac{ \mathrm{d}^3 k} {(2 \pi)^{3}} (f^{\vec{k}}_l- f^{\vec{k}}_n) \nonumber \\
    & \times& \frac{ \langle  n_{\vec{k}} |  \hat{v}_a   |  l_{\vec{k}} \rangle 
                    \langle  l_{\vec{k}} |  \hat{v}_b   |  m_{\vec{k}} \rangle
                    \langle  m_{\vec{k}} |  \hat{v}_c   |  n_{\vec{k}}  \rangle}
            {(E_{{\vec{k}}n}-E_{{\vec{k}}m}-i\Gamma)(E_{{\vec{k}}n}-E_{{\vec{k}}l} - \hbar \Omega- i\Gamma)} \bigg\}
\label{shift-eq}
\end{eqnarray}
where $\phi_{ab}$ is the phase difference between the driving field ${E}_a$ and ${E}_b$.
Here, $|n_{\vec{k}}\rangle$,  $E_{{\vec{k}}n}$, $m_0$, and $\hat{v}_b=\hat{p}/m_0$
 are electronic state, associated energy, free-electron mass and quasi-particle velocity operator along $b$-direction, respectively.
For circularly (linearly) polarized light, $\phi_{ab}$ becomes imaginary (real).   We note that $\phi_{ab}=i$ and $-i$, correspond to right and left circularly polarized light, respectively.  Hence, the  photocurrent along $c$-direction changes its sign  under the reversal of polarization of the light having electric fields along $a$ and $b$-direction \cite{zhang2018photogalvanic}. 
For example, one 
can consider the polarization vector $(0,1,i)$ and  $(0,1,-i)$ for right and left  circularly polarized light. Once the polarization of  the  circularly polarized light changes, 
the relative phases between the electric fields $E_y$ and $E_z$ also changes  such that $\phi_{yz}= i \to -i$ and eventually leading to the reversal in the direction for the photocurrent $J_x$.
The imaginary (real) part of the integral in Eq.~(\ref{shift-eq}) describes the CPGE  (shift current response) under circularly (linearly) polarized light \cite{sadhukhan2020first,pal2021bulk}.

We now discuss the photoconductivity formula as given in Eq.~(\ref{shift-eq}) in more detail. The formula is based on the three band transition where an additional virtual band is considered in addition to valence and conduction band.
The three band transitions are given by $n \rightarrow m \rightarrow l$ and $l \ne m$, whereas, two band transitions are given by $l = m$.   
It has been found that the two-band real transitions contribute much less in photocurrent as WN contributes maximally.
On the other hand, virtual transitions from the occupied Weyl to the empty Weyl
band via a third trivial band predominantly contribute
to the  high value of shift current and CPGE \cite{zhang2018photogalvanic}.  At the same time, we note that quantized \blue{CPGE} responses are observed in various model systems without inversion symmetry employing the two band  formula for CPGE that we discuss at the end of this section \cite{ le2020ab,de2017quantized,sadhukhan2020role}.

We below probe the behavior of photocurrent by plausible analytical argument.  At the outset,  we note that imaginary part of velocity numerator $N
= \langle  n_{\vec{k}} |  \hat{v}_a   |  l_{\vec{k}} \rangle 
\langle  l_{\vec{k}} |  \hat{v}_b   |  m_{\vec{k}} \rangle
\langle  m_{\vec{k}} |  \hat{v}_c   |  n_{\vec{k}}  \rangle$ would survive after the integral for  time reversal symmerty invariant non-magnetic WSM. Since the real part of the photoconductivity formula Eq.~(\ref{shift-eq}) gives the CPGE current,  the energy denominator $D=(E_{{\vec{k}}n}-E_{{\vec{k}}m}-i\Gamma)(E_{{\vec{k}}n}-E_{{\vec{k}}l} + \hbar \Omega- i\Gamma)$ has to be real under this circumstances.  The momentum integration in Eq.~(\ref{shift-eq}) acquires imaginary values for CPGE photoconductivity.  One can continue the calculation of photoconductivity by retaining the real part of
$D$ as ${\rm Re} [D]=
- \delta(E_{\vec{k}, n}-E_{\vec{k}, m})/\Gamma $.
We note that 
$\delta(E_{\vec{k}, n}-E_{\vec{k}, m})$ comes from the imaginary part of $(E_{{\vec{k}}n}-E_{{\vec{k}}m}-i\Gamma)^{-1}$ and the term $1/\Gamma$ arises from the second part $(E_{{\vec{k}}n}-E_{{\vec{k}}l} + \hbar \Omega- i\Gamma)^{-1}$ due to the selection rule $E_{{\vec{k}}n}=E_{{\vec{k}}l} - \hbar \Omega$. We know that CPGE receives the contribution from closed optically activated momentum surface as mediated by $\delta$-functions \cite{de2017quantized,sadhukhan2020role}.
The momentum integration of $N/D$ becomes imaginary for the CPGE photocurrent. Considering the time reversal symmetry invariant nature of the material and type of polarization, one can perform the the momentum integration by taking into account the appropriate terms only.  Then the photoconductivity can be rewritten in the following approximated form
\begin{eqnarray}
&&\sigma^c_{ab}(\omega)\nonumber
\\
&=&  \frac{e^3}{\omega^2} {\rm Re} \Bigg[ \phi_{ab}\sum_{\Omega=\pm \omega} \sum_{l, n, m} \int \frac{ \mathrm{d}^3 k} {(2 \pi)^{3}} f^{\vec{k}}_{nl}
\nonumber \\
    && \frac{ \langle  n_{\vec{k}} |  \hat{v}_a   |  l_{\vec{k}} \rangle 
                    \langle  l_{\vec{k}} |  \hat{v}_b   |  m_{\vec{k}} \rangle
                    \langle  m_{\vec{k}} |  \hat{v}_c   |  n_{\vec{k}}  \rangle}
            {(E_{{\vec{k}}n}-E_{{\vec{k}}m}-i\Gamma)(E_{{\vec{k}}n}-E_{{\vec{k}}l} - \hbar \Omega- i\Gamma)}   \Bigg] 
            \nonumber \\
 &\approx&
- \frac{e^3}{\omega^2}  \Bigg[ \sum_{\Omega=\pm \omega}
 \sum_{l, n, m} \int \frac{ \mathrm{d}^3 k} {(2 \pi)^{3}} f^{\vec{k}}_{nl} \delta(E_{\vec{k} n}-E_{\vec{k} m}) \nonumber \\ 
&&  \frac{ \langle  n_{\vec{k}} |  \hat{v}_a   |  l_{\vec{k}} \rangle 
                    \langle  l_{\vec{k}} |  \hat{v}_b   |  m_{\vec{k}} \rangle
                    \langle  m_{\vec{k}} |  \hat{v}_c   |  n_{\vec{k}}  \rangle}
                 {(E_{{\vec{k}}n}-E_{{\vec{k}}l} - \hbar \Omega- i\Gamma)}  \Bigg]
                  \nonumber\\
&\approx&
-\frac{e^3}{\omega^2}  \Bigg[ \sum_{\Omega=\pm \omega} \sum_{l, n} \int \frac{ \mathrm{d}^3 k} {(2 \pi)^{3}} f^{\vec{k}}_{nl} \frac  { { \langle  n_{\vec{k}} |  \hat{v}_a   |  l_{\vec{k}} \rangle 
                    \langle  l_{\vec{k}} |  \hat{v}_b   |  n_{\vec{k}} \rangle}
             {\langle  n_{\vec{k}} |  \hat{v}_c   |  n_{\vec{k}}}  \rangle } {(E_{{\vec{k}}n}-E_{{\vec{k}}l} - \hbar \Omega- i\Gamma)} \Bigg]
             \nonumber\\ 
&\approx&
-\frac{e^3}{ \hbar^2 \omega^2}  \Bigg[  \sum_{\Omega=\pm \omega} \sum_{l, n} \int \frac{k^{2} \mathrm{d} k_c d \Sigma}{(2 \pi)^{3}} f^{\vec{k}}_{nl}
     \frac{ E^2_{{\vec{k}},nl}  R^c_{{\vec{k}},nl} {\partial_{k_{c}} E_{{\vec{k}n} }} }
            {(E_{{\vec{k}}n}-E_{{\vec{k}}l} - \hbar \Omega- i\Gamma)}  \Bigg]
            \nonumber\\
&\approx&
-\frac{e^3}{\hbar^2 \omega^2}  \Bigg[ \sum_{\Omega=\pm \omega}
\sum_{l, n}\int \frac{ \mathrm{d} E_{{\vec{k}}} d \Sigma}{(2 \pi)^{3}}   \frac{E^2_{{\vec{k}},nl} k^{2} R^c_{{\vec{k}},n l} } {(E_{{\vec{k}}n}-E_{{\vec{k}}l} - \hbar \Omega- i\Gamma)}  \Bigg]
   \nonumber \\
&\approx& 
-  \frac{e^3 \tau }{\hbar \omega^2} \sum_{\Omega=\pm \omega}
\sum_{l, n}\int \frac{  d \Sigma}{(2 \pi)^{3}} \Omega^2  k^{2} R^c_{{\vec{k}},n l} \nonumber \\
&\approx&- \beta_0 \tau
\sum_{l, n}\int d S_{{\vec{k}},n l}  R_{{\vec{k}},n l}\nonumber \\ 
&\approx& i \beta_0 \tau
 \sum_{n} \int d \vec{S}_{n} \cdot \vec{\Omega}_{n} \nonumber \\            
&\approx& i C \beta_0 \tau
\nonumber \\
\label{three_band}
\end{eqnarray}
where $C= \sum_n C_{n}$ is Chern number summing over all the occupied bands and $\beta_0 = e^3/\hbar$.
  
Considering the relaxation time approximation for diffusive transport,  the relation between the broadening parameter $\Gamma$ and the quasiparticle lifetime $\tau$ is given by  $ \Gamma = \hbar/ \tau$ for metallic systems  \cite{zhang2018photogalvanic,ni2020giant,PhysRevB.88.104412,lundgren2014thermoelectric}.
In the above derivation, we first combine $\phi_{ab}\times (E_{{\vec{k}},n}-E_{{\vec{k}},m}-i\Gamma)^{-1}$ as $-\delta(E_{\vec{k}, n}-E_{\vec{k}, m})$ to retain the real part. Thereafter, we remove Re[...] considering the  fact that integration of $\frac{ \langle  n_{\vec{k}} |  \hat{v}_a   |  l_{\vec{k}} \rangle
\langle  l_{\vec{k}} |  \hat{v}_b| m_{\vec{k}} \rangle
\langle  m_{\vec{k}} |  \hat{v}_c   |  n_{\vec{k}}  \rangle} {(E_{{\vec{k}},n}-E_{{\vec{k}},l} - \hbar \Omega- i\Gamma)} $ yields quantized contributions to CPGE. To be precise, the terms ${\rm Im}[\langle  n_{\vec{k}} |  \hat{v}_a   |  l_{\vec{k}} \rangle
\langle  l_{\vec{k}} |  \hat{v}_b| m_{\vec{k}} \rangle
\langle  m_{\vec{k}} |  \hat{v}_c   |  n_{\vec{k}}  \rangle]$ and ${\rm Im}[(E_{{\vec{k}},n}-E_{{\vec{k}},l} - \hbar \Omega- i\Gamma)]$ combine together allowing the quantized contribution to CPGE within the closed 
optically activated momentum surface. We below demonstrate the technical steps with plausible arguments.

The topological charges of the activated WNs contribute to the quantization of photoconductivity. 
We consider the selection rule $E_{\vec{k}, n l}=E_{\vec{k}, l}-E_{\vec{k}, n} = \hbar \Omega$, owing to the optically activated momentum surface, and $f_{n l}^{\vec{k}}=f_{n}^{\vec{k}}-f_{l}^{\vec{k}}$ are difference between band energies and Fermi-Dirac
distributions, and $r^{a}_{\vec{k}, n l}=i\left\langle l_{\vec{k}}\left|\partial_{k_{a}}\right| n_{\vec{k}}\right\rangle = i\frac{ \hbar {\hat{v}}^a_{{\vec{k}},nl}}{E_{{\vec{k}},nl}}$. For $n \neq l$,  $r^{a}_{\vec{k}, n l}$ are the interband position matrix elements or off-diagonal Berry connection. For $n=l$, $r^{a}_{\vec{k}, n n}$ is the diagonal Berry connection. For the general case with $n\ne l$, $R^{c}_{\vec{k},nl}=\epsilon_{a b c}r_{\vec{k}, n l}^{a} r_{\vec{k}, l n}^{b}$.  We note that $\langle  n_{\vec{k}} |  \hat{v}_a   |  l_{\vec{k}} \rangle                     \langle  l_{\vec{k}} |  \hat{v}_b   |  n_{\vec{k}} \rangle $, coming from the imaginary numerator as described above, can be  written as $E^2_{{\vec{k}},nl}  R^c_{{\vec{k}},nl}$. On the other hand, the energy integration $\int  \mathrm{d} E_{{\vec{k}}} [\frac{E^2_{{\vec{k}},nl}} {(E_{{\vec{k}},n}-E_{{\vec{k}},l} - \hbar \Omega- i\Gamma)}]$, assuming the imaginary part of denominator contributes to CPGE,  reduces to $\frac{\hbar^2 \Omega^2}{\Gamma}$. 
Upon judiciously implementing all the above mathematical steps, one can bring down the quantity 
$\int  d k_c [\frac{ \langle  n_{\vec{k}} |  \hat{v}_a   |  l_{\vec{k}} \rangle
\langle  l_{\vec{k}} |  \hat{v}_b |n_{\vec{k}} \rangle
\partial_{k_{c}} E_{{\vec{k},n} }
} {(E_{{\vec{k}},n}-E_{{\vec{k}},l} - \hbar \Omega- i\Gamma)} ]$ to the following form $\frac{ \hbar^2 \Omega^2  R^c_{{\vec{k}},n l} } {\Gamma}$. 
The relation between $\vec{R}_{\vec{k},n l}$ and Berry curvature
is given by $\vec{\Omega}_{\vec{k},n}=i \sum_{l \neq n} \vec{R}_{\vec{k},n l}$. 
Here $\vec{S}_n$ is a closed surface of band $n$ enclosing
the degenerate points. 
For a given frequency $\omega$,  the delta function and Fermi-Dirac distributions select a surface $\vec{S}_{nl}$
in the $\vec{k}$ space where  $d\vec{S} = k^2 d\Sigma$ denotes the surface element oriented 
normal to $\vec{S}$ where $d\Sigma$ is the differential solid angle.  Therefore,  the CPGE current is essentially the Berry flux penetrating through surface $\vec{S}$.    We note that 
trace of CPGE tensor ${\rm Tr}[\beta]/i\beta_0$ is found to be quantized \cite{de2017quantized,sadhukhan2020role,le2020ab}.


\par We now derive the two-band formula for the photocurrent ${ \tilde\sigma}^c_{ab}(\omega)$ from three-band formula by accounting the direct optical transition from $\ket{l_{\vec{k}} }$ to $\ket{n_{\vec{k}} }$ as given in Eq.~(\ref{shift-eq}).
We consider $n=m$ and 
$(E_{{\vec{k}},n}-E_{{\vec{k}},m}-i\Gamma)^{-1}=  i/\Gamma = i \tau/ \hbar $. {One can allow a broadening $\Gamma$ around a given energy level  $E_{{\vec{k}},n}$. This comes into play when any two energy levels become equal to each other within the quantum limit. }  Using $E_{{\vec{k}},nl} =\pm \hbar \omega$, can write 
$r^{a}_{\vec{k}, n l}= i \frac{\hbar{\hat{v}}^a_{{\vec{k}},nl}}{E_{{\vec{k}},nl}} = i \frac{{\hat{v}}^a_{{\vec{k}},nl}}{\omega}$. 
The photoconductivity 
as reduced from   Eq.~(\ref{shift-eq}), is given below  \cite{de2017quantized,ahn20,flicker2018chiral}

{
\begin{eqnarray}
{\tilde \sigma}^c_{ab}(\omega) = \frac{ e^3 \tau}{  \hbar} \epsilon_{bfg} 
\sum_{l,n} \int_{BZ} \frac{ \mathrm{d}^3 k} {(2 \pi)^{3}} f^{\vec{k}}_{nl}
{\Delta}^a_{{\vec{k}},nl}
{r}^f_{{\vec{k}},nl}
{r}^g_{{\vec{k}},ln} 
\delta(\hbar \omega-E_{\vec{k}, n l})   \nonumber \\
\label{cpge-2band-eq}
\end{eqnarray}}
{with $\hat{v}^f_{\vec{k},nl}= \langle  n_{\vec{k}} |  \hat{v}_f   |  l_{\vec{k}} \rangle$, $E_{\vec{k},nl}= E_{{\vec{k}},n}-E_{{\vec{k}},l}$ and $\Delta^a_{\vec{k},nl}=\hat{v}^a_{\vec{k},nn} - \hat{v}^a_{\vec{k},ll}$.} {Notice that $\epsilon_{bfg}$ represents the Levi-Civita symbol and two band formula for photoconductivity is usually referred to as ${\tilde \sigma}_{ab}(\omega)$ in the literature.}
{Here we have already considered $\phi_{ab}=i$ while computing the two band formula Eq.~(\ref{cpge-2band-eq}) from three band formula Eq.~(\ref{shift-eq}). 
One there has to be careful with the summation 
$ \sum_{l, n} f^{\vec{k}}_{nl}
\hat{v}^a_{\vec{k},nl}  \hat{v}^b_{\vec{k},ln} \hat{v}^c_{\vec{k},nn}
$. We below discuss the analytical reduction from three band formula (Eq.~(\ref{shift-eq})) to two band with plausible argument. }

{The summation is over the repeated indices, and $l(n) \to n (l)$, one can find 
$f^{\vec{k}}_{nl}
\hat{v}^a_{\vec{k},nl}  \hat{v}^b_{\vec{k},ln} \hat{v}^c_{\vec{k},nn} \to 
f^{\vec{k}}_{ln}
\hat{v}^a_{\vec{k},ln}  \hat{v}^b_{\vec{k},nl} \hat{v}^c_{\vec{k},ll}$.
The term $\hat{v}^a_{\vec{k},nl}  \hat{v}^b_{\vec{k},ln}$ can be written in an  anti-symmetric way $A(a,b)=\frac{1}{2}(\hat{v}^a_{\vec{k},nl}  \hat{v}^b_{\vec{k},ln} - \hat{v}^a_{\vec{k},ln}  \hat{v}^b_{\vec{k},nl})$ such that $A(a,b)$ reverse its sign under the reversal of polarization $A(a,b)=-A(b,a)$ with $a(b) \to b (a)$. 
Now, we can decompose the summation with $l(n) \to n (l)$ and find $\sum_{n, l} \frac{1}{2}\Big[ f^{\vec{k}}_{nl} A(a,b) \hat{v}^c_{\vec{k},nn} + f^{\vec{k}}_{ln} A(a,b)
\hat{v}^c_{\vec{k},ll} \Big] =  
\sum_{n, l} \Big[ f^{\vec{k}}_{nl} \frac{(\hat{v}^a_{\vec{k},nl}  \hat{v}^b_{\vec{k},ln} - \hat{v}^a_{\vec{k},ln}  \hat{v}^b_{\vec{k},nl})}{2} \Delta^c_{\vec{k},nl}
\Big] $ with ${\Delta}^c_{\vec{k},nl} = {\partial_{k_{c}} E_{{\vec{k},nn} }} - {\partial_{k_{c}} E_{{\vec{k},ll} }}=\hat{v}^c_{\vec{k},nn} - \hat{v}^c_{\vec{k},ll}$  and $f^{\vec{k}}_{nl} = - f^{\vec{k}}_{ln}$. As a result, the two-band
photoconductivity formula looks like 
 \begin{eqnarray}
{\tilde \sigma}^c_{ab}(\omega) &=&  \frac{e^3\tau}{\omega^2\hbar} \sum_{\Omega=\pm \omega} \sum_{l, n} \int \frac{ \mathrm{d}^3 k} {(2 \pi)^{3}} f^{\vec{k}}_{nl}
\frac{1}{2}(\hat{v}^a_{\vec{k},nl}  \hat{v}^b_{\vec{k},ln} - \hat{v}^a_{\vec{k},ln}  \hat{v}^b_{\vec{k},nl}) \nonumber \\
&& \times 
\Delta^c_{\vec{k},nl}
\delta(E_{\vec{k},nl}- \hbar \Omega)    
\end{eqnarray}
In order to cast the above expression in terms of 
$r^{a,b}_{\vec{k},nl}$, one has to use $ \hat{v}^{a,b}_{\vec{k},nl} = -i r^{a,b}_{\vec{k},nl} E_{{\vec{k}},nl}/\hbar$ with the selection rule 
$E_{{\vec{k}},nl}= \pm \hbar \omega$, provided $E_{{\vec{k}},nl}= \hbar \Omega$, in accordance with the delta function $\delta(E_{{\vec{k}},nl}- \hbar \omega)$. Therefore, by replacing $\delta(E_{{\vec{k}},nl}- \hbar \Omega)$ with $\delta(E_{{\vec{k}},nl}- \hbar \omega)$,
and absorbing the 
summation over $\Omega=\pm \omega$,
the photoconductivity is found to be 
\begin{eqnarray}
{\tilde \sigma}^c_{ab}(\omega)
 = \frac{ e^3 \tau}{ \hbar} 
 \sum_{l,n} \int_{BZ} \frac{ \mathrm{d}^3 k} {(2 \pi)^{3}} f^{\vec{k}}_{nl}[ r_{\vec{k}, n l}^{a} r_{\vec{k}, l n}^{b}]  {\Delta}^c_{ln}  {\delta(E_{\vec{k}, nl}-\hbar \omega)} \nonumber \\
\label{cpge-2band-eq2}
\end{eqnarray}
with $[r_{\vec{k}, n l}^{a}, r_{\vec{k}, l n}^{b} ]=(r^a_{\vec{k},nl}  r^b_{\vec{k},ln} - r^a_{\vec{k},ln}  r^b_{\vec{k},nl})/2$. One can thus obtain two band CPGE formula Eq.~(\ref{cpge-2band-eq}), starting from the three
band approach, where the reversal of polarization  appropriately leads to a reversal in sign ${\tilde \sigma}^c_{ab}(\omega)=-{\tilde \sigma}^c_{ba}(\omega)$ owing to the above commutator like relation $[r_{\vec{k}, n l}^{a}, r_{\vec{k}, l n}^{b} ]= -[r_{\vec{k}, n l}^{b}, r_{\vec{k}, l n}^{a} ]$.}
{We note that Eqs.~(\ref{cpge-2band-eq2}) and (\ref{cpge-2band-eq})  are connected by appropriate re-arrangement of indices in Levi-Civita symbol.}
We  below examine the quantization in the photoconductivity from the three and two band approaches.


\begin{figure}[ht]
\includegraphics[width=0.47\textwidth,angle=0]{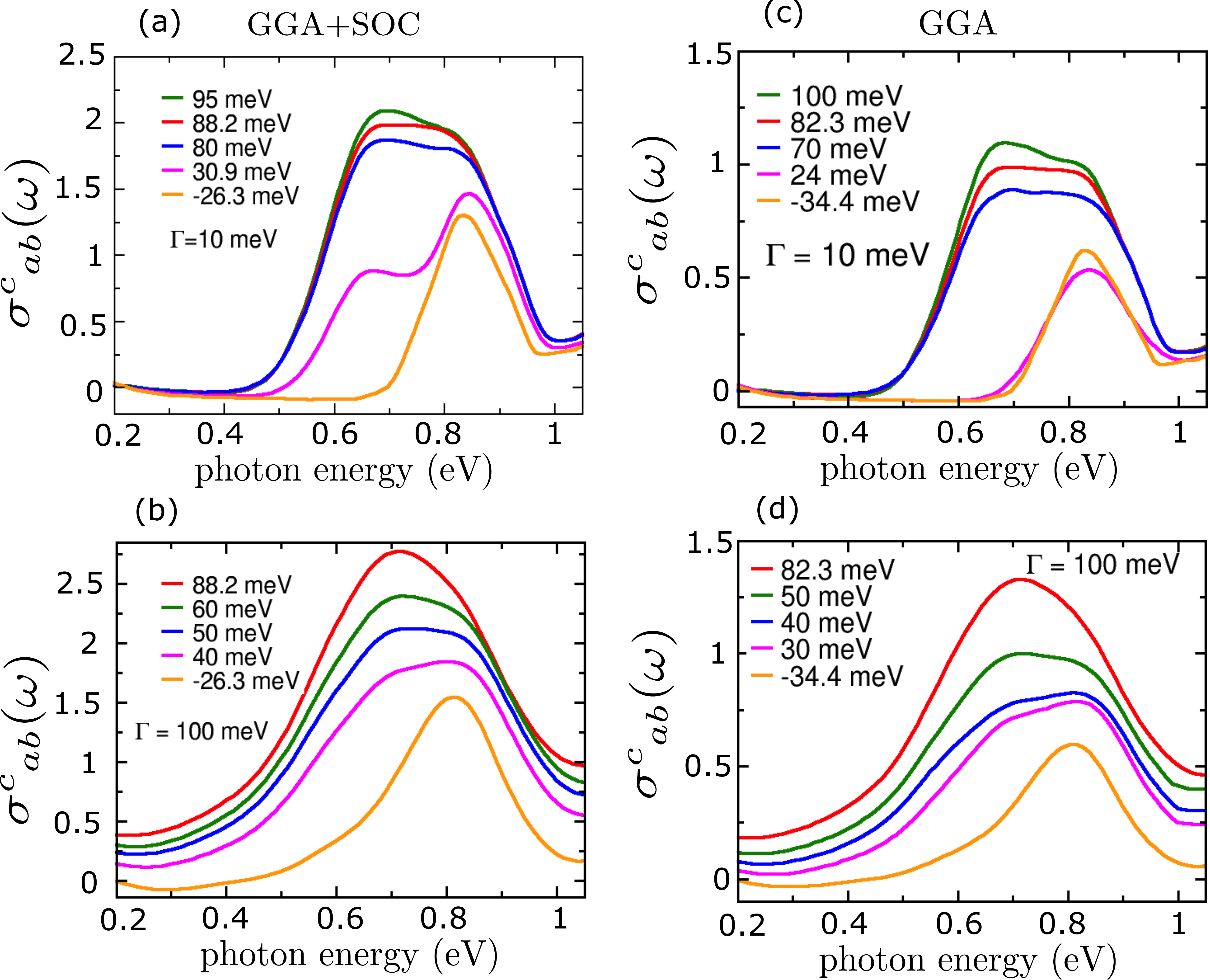}
\caption{ CPGE, computed from Eq.~(\ref{shift-eq}), with  GGA (a)-(b) and  GGA+SOC (c)-(d) for broadening parameter $\Gamma = 10$ meV  and $\Gamma = 100$ meV respectively.  We plot $\sigma_{ab}^c = \sigma_{ab}^c/i$ in all the figures.  The different colored lines correspond to different chemical potentials as denoted in the figures.  The red (orange) line in each figures represents the CPGE response when $E_f$ is set at energy associated with WN of negative (positive) chirality.
\label{fig:cpge}} 
\end{figure}

According to symmetry classification of the material, the non zero elements of the CPGE tensor for SrSi$_2$ are $\sigma_{yz}^x$,  $\sigma_{zx}^y$, $\sigma_{xy}^z$.   To be precise,  $\sigma_{yz}^x = - \sigma_{zx}^y = \sigma_{xy}^z $. The relaxation time plays important role in determining the quantization of  CPGE.  Therefore it is indeed very important to have  correct estimation of relaxation time to obtain 
reliable CPGE.  Typically,  the hot-electron scattering time for metallic systems $\approx fs $ \cite{ni2020linear} which corresponds to a broadening parameter $\Gamma = \hbar /  \tau   = 10 - 100$  meV.  The effect of temperature and impurity scattering on photocurrent generation are taken into account by this broadening parameter $\Gamma$.  The BZ was sampled by ${\bm k}$-mesh with 250 $\times$ 250  $\times$ 250 to compute the CPGE current.

\par Figure \ref{fig:cpge} (a)-(b) and (c)-(d) shows the CPGE response for $ \Gamma = $10,  100  meV with and without SOC respectively. We find the  chiral chemical potential $\mu_{ch} =  E_{V_2}-E_{V_1}= 116.7$ ($= E_{W_2}-E_{W_1}= 114.5$) meV for GGA  (GGA+SOC) calculations. The magnitude of topological charge   changes whether SOC is excluded or included in the DFT calculation. This is clearly reflected in the  
quantization of  photoconductivity while studied it in presence and absence of SOC i.e., photoconductivity is  found to be quantized around the value $1$ ($2$) for  GGA (GGA+SOC) calculations. 
The magnitude of the quantization  is governed by the topological charge of the activated  WN.   This is in accordance with the theoretical conjecture as discussed in the earlier section. 
With inclusion of SOC,  we further observe the quantization within the energy $\approx$ $0.6 <  \omega <  0.9$ eV is very prominent compared to the GGA results
when the Fermi level is kept near one of the WNs at $E_f=E_{W_2}=88.2$ meV for $\Gamma = 10$ meV as shown in Fig.\ref{fig:cpge} (a).  
The  $\sigma_{yz}^x$ and $\sigma_{zx}^z$  components of CPGE response follow the same sign of the activated WN at $E_{W_2}=88.2$ meV ($n=-2$) whereas the $\sigma_{yz}^y$ reverses its sign.

The frequency windows for quantization is given by $2|E'_{W_1}|<\omega < 2 |E'_{W_2}|$ with $E'_{W_1, W_2}=E_{W_1,W_2}-E_f$. For, $E_f=88.2$ meV,  the frequency window for quantization becomes $0 <  \omega <  0.3$ eV that does not match with our numerical findings. However, 
the extent of frequency interval within which quantization occurs i.e., $0.3$ eV, predicted from analytical analysis, matches well with the numerical finding. 
We find that the quantization starts around $\omega \simeq 0.6$ eV which  can be due to the presence of other bands lying in the vicinity of  WN energies $E_{W_2}$ and $E_{W_1}$ and their non-linear dispersive nature. The quantization within a given frequency window can only be predicted by the low-energy mode. On the other hand, in real material there exist a variety of non-linear and non-trivial effects causing the deviation from the exact quantization.

We observe a deviation from quantization for $E_f=95.0$, $80.0$ meV away from the WN energies. Interestingly, when we keep the Fermi level near other WN (with topological charge $n=2$) energy at $E_{W_2}=-26.3$ meV,  a non-zero but non quantized response of CPGE is observed. 
It is expected that WN with $n=+2$ will 
contribute to the photoconductivity.
We observe qualitatively the same behavior  for $E_f = 30.9 $ meV,   $E'_{W_1}=E'_{W_2}$  i.e., the Fermi level is exactly at midway between two WN energies. These results are qualitatively different from a time reversal symmetry broken WSMs where CPGE responses are quantized to two opposite values depending on the topological charge of the activated WN at a given Fermi level \cite{de2017quantized}. On the other hand,  CPGE response for  time reversal symmetry invariant WSM is found to be different from  time reversal symmetry broken WSM as the Berry curvature and velocity exhibit non-trivial behavior over BZ. We here find similar agreement in CPGE response for the non-magnetic material SrSi$_2$.  We obtain similar findings with  GGA  calculations as shown in Fig.~\ref{fig:cpge} (c) and (d). 
We note that with increasing broadening parameter $\Gamma$, the quantization of photoconductivity is lost and it can acquire values more than the magnitude of topological charge (see Fig.~\ref{fig:cpge} (b) and (d)). This is due to the fact that more states contribute to the photoconductivity around a given Fermi level. Therefore, photoconductivity exhibits quantization only when $\Gamma$ is less than the energy difference between two consecutive bands around a given Fermi energy in the vicinity of WN energy.

\begin{figure}[ht]
\includegraphics[width=0.5\textwidth,angle=0]{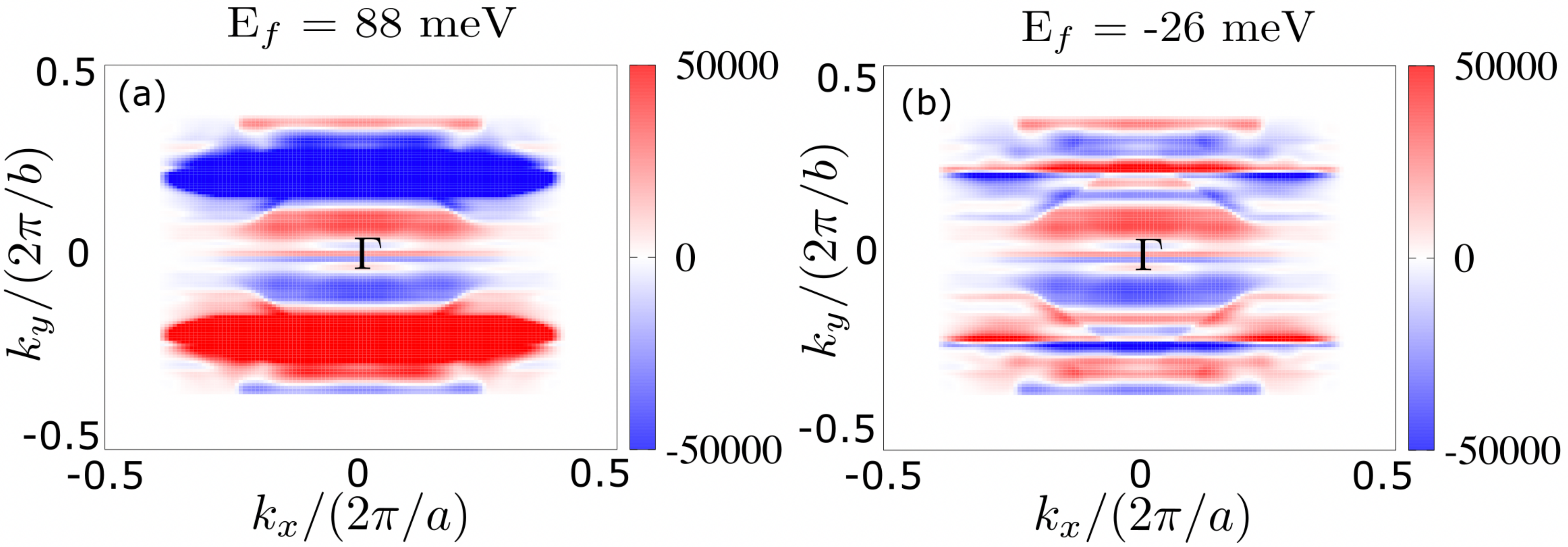}
\caption{Momentum resolved $k_z$-averaged
  photoconductivity considering $\omega=0.75$ eV for two energies at (a)  $E_f=88.2$ meV and (b)$E_f=-26.3$ meV respectively associated with WNs of opposite chirality .
\label{moment-avg}} 
\end{figure}

To understand the distinct behavior of photoconductivity around two  WNs in different energies $E_f=88.2,  -26.3$ meV,   we analyze the momentum resolved structure of CPGE conductivity as shown in Fig.~\ref{moment-avg} (a) and (b). We evaluate Eq.~(\ref{shift-eq}) as a function of $k_x$ and $k_y$, integrating over $k_z$,
at a given value of $\omega=75.0$ meV for which photoconductivity becomes quantized. We find that  the structure of 
the momentum resolved photoconductivity for $E_f=88.2$ meV is substantially different from that of the for $E_f=-26.3$ meV. These two 
different momentum distributions at 
two WNs with opposite chirality clearly refers to the fact that CPGE will not be opposite of each other for these two WNs.  This is due to the fact that the
Berry curvature and velocity, essentially  determining the response, 
behave differently at the two WNs with opposite chirality  for time reversal symmetric systems \cite{sadhukhan2020role}.

\begin{figure}[ht]
\includegraphics[width=0.5\textwidth,angle=0]{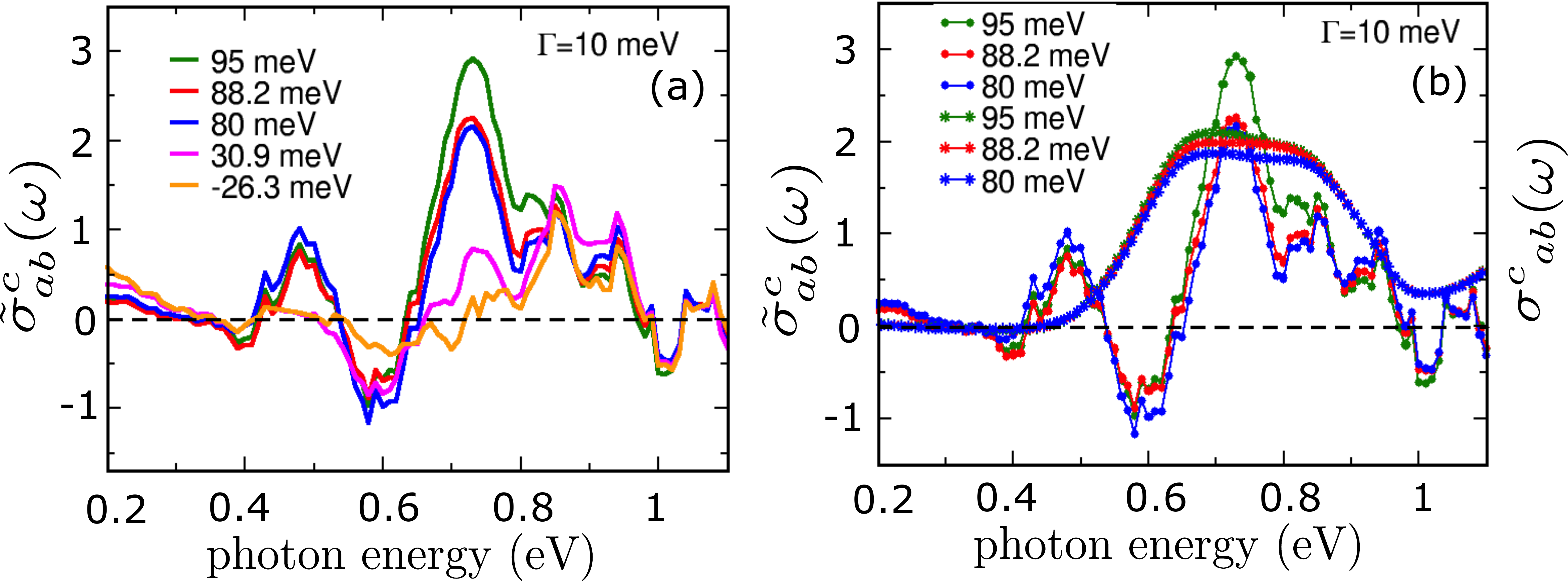}
\caption{ (a) CPGE  from 2-band  approach Eq.~(\ref{cpge-2band-eq}) with SOC using the {Lorentzian}  broadening parameter $\Gamma = 10$ meV { while numerically implementing  $\delta(\hbar \omega-E_{\vec{k}, n l})$ \cite{ni2020giant}.}  (b) Comparison of CPGE response from 2-band (marked by circle) and 3-band (marked by star) approach using the same broadening parameter respectively.  We plot ${\tilde \sigma}_{ab}^c={\tilde \sigma}_{ab}^c/i$ in all the figures. The different colored lines correspond to different chemical potentials as designated in the figures.
\label{fig:cpge-compare}} 
\end{figure}

We now compare our findings, followed by three band formula (Eq.~(\ref{shift-eq})), with the two band formula (Eq.~(\ref{cpge-2band-eq})) as shown in Fig.~\ref{fig:cpge-compare} (a) and (b) , considering  GGA+SOC calculations. It is evident that there is no quantization  observed from two band formula suggesting the important role played by the virtual transition channel as incorporated in the three band formula. However,  the CPGE, obtained from Eq.~(\ref{cpge-2band-eq}) becomes qualitatively pronounced  within the same frequency window for which quantization observed following Eq.~(\ref{shift-eq}). On the other hand, we commonly observe that CPGE are not opposite of each other for $E_f=88.2$ meV and $E_f=-26.3$ meV. Therefore, 
in order to predict the correct behavior as far as the quantization is concerned, the three band formula is found to be useful in examining the optical responses of semimetals \cite{sadhukhan2020first}.

We here compare with the quantized CPGE response for multifold fermion transition metal mono-silicides, with  higher spin degrees of freedom,  where multiple bands show linear band crossing at the degenrate points \cite{flicker2018chiral}. By contrast, SrSi$_2$, with an effective spin-$1/2$ degrees of freedom, exhibits non-linear band crossing at the WNs in presence of SOC. However, in both the materials, the inversion and mirror symmetries are broken resulting in non-degenerate WNs with finite chiral chemical potential. The quantized response in CPGE is a consequence of that while non-centrosymmetric WSM TaAs with degenrate WNs shows non-quantized CPGE response \cite{zhang2018photogalvanic}. Therefore, the linear band dispersion along at least one momentum around the gap closing point , noticed for all of the above materials, might not be directly responsible for the quantized response unless there exist finite chiral chemical potential. Having said that, we note that the 
magnitude of quantization depends on the underlying topological structure of these materials.

In the case of multifold fermions,
to be precise, the  higher
spin degrees of freedom essentially leads to the higher order topological charge. The optically activated momentum surface might
include more than a single band. This equivalently results in the quantized CPGE
trace coming from the Berry curvatures associated with more than a single band. Therefore, Chern numbers, associated with the
various activated topological bands (i.e., within the optically activated momentum
surface) are essentially responsible for the high value of quantization for CPGE trace in multifold fermions \cite{flicker2018chiral}. On the other hand, the quadratic energy dispersion of SrSi$_2$ with SOC (see Fig.~\ref{fig:band} (a)) imprints its signature in the quantized CPGE trace via the topological charge even though it does not have more than twofold degeneracy or higher spin degrees of freedom as found in multifold fermion. The Berry curvature of a single band, within the optically activated momentum
surface, would contribute to the quantized CPGE trace in double WSM SrSi$_2$. This is in contrast to the multifold fermions where Berry curvatures, coming from different bands can add up to give rise the high value of quantized CPGE trace.

{The quantization in multifold fermions is observed following the two band formula (Eq.~(\ref{cpge-2band-eq})) \cite{flicker2018chiral} while the quantization in present case of double WSM is based on the three band formula (Eq.~(\ref{shift-eq})). This is apparently very much intriguing and it requires extensive future investigations. However, the deviation from quantization following two band formula in the present case, is not related to the broadening parameter as this parameter appears both in two band as well as three band formulas.  It might be that the three band approach is able mimic the effect of anisotropic non-linear bands in double WSM more vividly than the two band approach. Therefore, it would be really an interesting future direction to systematically explore the photoconductivity in various other suitable materials.}

\section{Conclusions}
\label{conclusion}

\par To summarize, considering SrSi$_2$ as a non-centrosymmetric and non-magnetic double-WSM we first study the structural handedness to show enantiomeric nature of the material. This manifests itself through the  SFS profile in $(001)$ and $(00\bar 1)$ surface. 
The large $S$ [inverted $S$]  shaped Fermi arcs  in the high symmetry direction $\Gamma-\ X$ for   $(001)$ [$(00\bar 1)$] surface   referring to the mirror symmetric but not superimposable nature of SFS between  $(001)$ [$(00\bar 1)$] surfaces. On the other hand, 
bulk Fermi surface depicts the emergence of electron and hole pocket suggesting type-I nature of the WSM. We perform GGA and GGA+SOC calculations to examine the topological properties of the material  from surface energy spectrum where we find  singlet and doublet chiral Fermi arc, respectively. Unlike the SFS, the surface energy dispersion  
between $(001)$ and $(00\bar 1)$ surfaces are mirror symmetric as well as superimposable with respect to each other. The enantiomeric property of the material thus leads to a chiral SFS and an achiral surface energy dispersion.

Having investigated the connection between structural handedness and chirality of surface Fermi surface profiles, we now study the CPGE as the  there exists a substantial gap between the energies of WNs with opposite chirality. The CPGE shows a quantized plateau depending on the topological charge of the activated WN when Fermi level is set only around the WN with higher energy. This is markedly different from the time reversal symmetry breaking WSM where CPGE is quantized to two exactly  opposite   plateau,
depending on the topological charge of the respective  activated WN, for two different energies.  
We analyze the momentum resolved structure of CPGE  at the two different WN energies to strengthen our findings. We additionally compare our results with the two-band formula to show that three-band formula captures the virtual transitions from Weyl bands to trivial bands in determining the accurate  photocurrent responses.


\section{Acknowledgments}
BS thanks Jeroen van den Brink,  Yang Zhang,  Yan Sun for discussions and Ulrike Nitzsche for technical assistance.  The calculations were carried out in the IFW cluster. {We sincerely acknowledge the anonymous referees for their constructive suggestons.}

\bibliography{srsi2}


\end{document}